%% file: main.tex
\documentclass[
    reprint,
    amsmath,
    amssymb,
    aps,
    prb,
    superscriptaddress
]{revtex4-2}

\usepackage{graphicx}
\usepackage{dcolumn}
\usepackage{bm}
\usepackage{booktabs}

\usepackage{siunitx}
\DeclareSIUnit{\cps}{cps}

\usepackage[english]{babel}
\usepackage[utf8]{inputenc}
\DeclareUnicodeCharacter{2212}{\ensuremath{-}}

\AtBeginDocument{%
  \let\oldbibliography\bibliography
  \renewcommand{\bibliography}[1]{%
    \makeatletter\def\selectlanguage##1{}\makeatother
    \oldbibliography{#1}%
  }%
}

\usepackage[
    colorlinks=true,
    linkcolor=blue,
    citecolor=blue,
    urlcolor=blue
]{hyperref}

\begin{document}

\title{High-cooperativity coupling and spin-resolved extinction of tin-vacancy centers in a diamond-like microcavity}

\author{Kerim Köster}
\affiliation{Physikalisches Institut (PHI), Karlsruhe Institute of Technology (KIT), Wolfgang-Gaede-Straße 1, 76131 Karlsruhe, Germany}

\author{András Laukó}
\affiliation{Physikalisches Institut (PHI), Karlsruhe Institute of Technology (KIT), Wolfgang-Gaede-Straße 1, 76131 Karlsruhe, Germany}

\author{Federico Rapisarda}
\affiliation{Physikalisches Institut (PHI), Karlsruhe Institute of Technology (KIT), Wolfgang-Gaede-Straße 1, 76131 Karlsruhe, Germany}

\author{Philipp Graßhoff}
\affiliation{Institute of Nanostructure Technologies and Analytics (INA), Center for Interdisciplinary Nanostructure Science and Technology (CINSaT), University of Kassel, Heinrich-Plett-Straße 40, 34132 Kassel, Germany}

\author{Vladislav Bushmakin}
\affiliation{3rd Institute of Physics, University of Stuttgart, Allmandring 13, 70569 Stuttgart, Germany}

\author{Jens Fuhrmann}
\affiliation{Institute for Quantum Optics, Ulm University, Albert-Einstein-Allee 11, 89081 Ulm, Germany}

\author{Ou Wang}
\affiliation{Institute for Quantum Optics, Ulm University, Albert-Einstein-Allee 11, 89081 Ulm, Germany}

\author{Dominic Reinhardt}
\affiliation{Division of Applied Quantum Systems, Felix-Bloch-Institute for Solid State Physics, University of Leipzig, 04103 Leipzig, Germany}

\author{Doğuşcan Ahiboz} 
\affiliation{Ferdinand-Braun-Institut (FBH), Gustav-Kirchhoff-Str. 4, 12489 Berlin, Germany}
\affiliation{Department of Physics, Humboldt-Universität zu Berlin, Newtonstr. 15, 12489 Berlin, Germany}

\author{Peter Knittel}
\affiliation{Fraunhofer Institute for Applied Solid State Physics, Tullastr. 72, 79108 Freiburg, Germany}

\author{Thomas Hümmer}
\affiliation{Qlibri GmbH, Maistr. 67, 80337, München, Germany}

\author{Wolfgang Wernsdorfer}
\affiliation{Physikalisches Institut (PHI), Karlsruhe Institute of Technology (KIT), Wolfgang-Gaede-Straße 1, 76131 Karlsruhe, Germany}
\affiliation{Institute for Quantum Materials and Technologies (IQMT), Karlsruhe Institute of Technology (KIT), Herrmann-von-Helmholtz Platz 1, 76344, Eggenstein-Leopoldshafen, Germany}

\author{Jörg Wrachtrup}
\affiliation{3rd Institute of Physics, University of Stuttgart, Allmandring 13, 70569 Stuttgart, Germany}

\author{Fedor Jelezko}
\affiliation{Institute for Quantum Optics, Ulm University, Albert-Einstein-Allee 11, 89081 Ulm, Germany}

\author{Tommaso Pregnolato}
\author{Tim Schröder}
\affiliation{Ferdinand-Braun-Institut (FBH), Gustav-Kirchhoff-Str. 4, 12489 Berlin, Germany}
\affiliation{Department of Physics, Humboldt-Universität zu Berlin, Newtonstr. 15, 12489 Berlin, Germany}

\author{Jan Meijer}
\affiliation{Division of Applied Quantum Systems, Felix-Bloch-Institute for Solid State Physics, University of Leipzig, 04103 Leipzig, Germany}

\author{Cyril Popov}
\affiliation{Institute of Nanostructure Technologies and Analytics (INA), Center for Interdisciplinary Nanostructure Science and Technology (CINSaT), University of Kassel, Heinrich-Plett-Straße 40, 34132 Kassel, Germany}

\author{David Hunger}
\email{david.hunger@kit.edu}
\affiliation{Physikalisches Institut (PHI), Karlsruhe Institute of Technology (KIT), Wolfgang-Gaede-Straße 1, 76131 Karlsruhe, Germany}
\affiliation{Institute for Quantum Materials and Technologies (IQMT), Karlsruhe Institute of Technology (KIT), Herrmann-von-Helmholtz Platz 1, 76344, Eggenstein-Leopoldshafen, Germany}

\date{\today}

\begin{abstract}
The tin-vacancy (SnV) center in diamond is a promising spin–photon interface for quantum networks, combining favorable optical properties with spin coherence above \SI{1}{\kelvin}. Unfolding the full potential requires cavity enhancement to increase photon–emitter coupling efficiency. Here, we demonstrate cavity-enhanced light–matter coupling of SnV centers in a fully tunable Fabry–P\'erot microcavity operating at temperatures down to \SI{1}{\kelvin} with in-situ magnetic field control. We access the diamond-like regime of hybrid cavity modes through integration of low-roughness diamond membranes, where the field is concentrated inside the diamond and Purcell enhancement is maximized. Diamond-like modes deliver a more than two-fold increase in the effective Purcell factor over air-like modes, reaching $C_0 = 4.1(1)$ compared to $C_0 = 1.85(5)$ in the air-like case, while simultaneously relaxing mechanical stability requirements. Resonant probing reveals coherent cavity–emitter coupling with 96\% extinction contrast and a coherent cooperativity of $C = 4.0(14)$. By applying a magnetic field, we further achieve spin-resolved cavity extinction, observing spin-selective optical transitions with a contrast of $\mathcal{C}_\mathrm{spin} = 0.91$. These results establish SnV centers in diamond coupled to open Fabry–P\'erot microcavities as a promising platform for efficient spin–photon interfaces.
\end{abstract}

\maketitle

\input{Chapters/1_Introduction}
\input{Chapters/2_Setup}
\input{Chapters/3_HybridCavity}
\input{Chapters/4_Purcell}
\input{Chapters/5_Extinction}
\input{Chapters/6_Zeeman}
\input{Chapters/7_Conclusion}
\begin{acknowledgments}
This work was partly supported by the German Federal Ministry of Research, Technology and Space (Bundesministerium für Forschung, Technologie und Raumfahrt, BMFTR) within the projects QR.N (Contracts No. 16KIS2186, No. 16KIS2204, No. 16KIS2180K, and No. 16KIS2185), QR.X (Contracts No. 16KISQ004, No. 16KISQ001K, and No. KIS6QK4001), SPINNING (Contract No. 13N16211), the European Union’s Horizon Europe research and innovation programme under grant agreement No 11154311, the Max Planck School of Photonics (MPSP), and the Karlsruhe School of Optics and Photonics (KSOP).
We thank the cleanroom team at the University of Kassel for the fabrication of diamond membranes, Qinu GmbH for technical support with the cryostat, and Qlibri GmbH for technical support with the microcavity stages.
\end{acknowledgments}
\appendix
\input{Chapters/8_Appendix}
\clearpage
\bibliography{references}

\end{document}

%% file: Chapters/1_Introduction.tex
\section{Introduction}

Quantum networks capable of distributing entanglement between remote nodes underpin key applications in quantum communication~\cite{kimble_quantum_2008,wehner_quantum_2018}, distributed sensing~\cite{gottesman_longer-baseline_2012,komar_quantum_2014,guo_distributed_2020}, and modular quantum computing~\cite{monroe_large-scale_2014,nickerson_freely_2014}. Realizing such networks requires nodes that combine long-lived spin memories with efficient optical interfaces for photon-mediated entanglement generation. The most advanced demonstration to date is a three-node quantum network based on nitrogen-vacancy (NV) centers in diamond, whose optically addressable electronic spin coupled to long-lived nuclear-spin memories enables multiple networking primitives within a single platform~\cite{pompili_realization_2021}. Yet scaling beyond this few-node demonstration is constrained by the low fraction of coherently emitted NV photons~\cite{faraon_coupling_2012} and the resulting low heralded entanglement rates~\cite{bernien_heralded_2013,hensen_loophole-free_2015}. Addressing this requires both photonic cavities that enhance light-matter coupling via the Purcell effect~\cite{reiserer_cavity-based_2015,janitz_cavity_2020} and emitter systems with intrinsically higher yield of indistinguishable photons.

Group-IV color centers in diamond fulfill the latter requirement through their inversion symmetry, which suppresses first-order Stark shift and thus reduces spectral diffusion, and yields a larger Debye--Waller factor~\cite{bradac_quantum_2019,ruf_quantum_2021}. Silicon-vacancy centers integrated into nanophotonic cavities have demonstrated near-unity light--matter coupling efficiencies and cooperativities exceeding 100~\cite{bhaskar_experimental_2020,riedel_scalable_2025}. This has enabled entanglement generation across a metropolitan-scale network, albeit at millikelvin temperatures or in highly strained environments required to preserve spin coherence~\cite{stas_robust_2022,knaut_entanglement_2023}. In contrast, heavier group-IV defects - in particular the tin-vacancy (SnV) center - combine favorable optical properties with operation above 1~K and long-lived electron- and nuclear-spin qubits, with microwave-addressable spin control~\cite{guo_microwave-based_2023,rosenthal_microwave_2023,karapatzakis_microwave_2024,resch_high-fidelity_2026, beukers_control_2025}, making them attractive candidates for spin--photon interfaces with relaxed cryogenic requirements, recently demonstrated via waveguide-integrated remote entanglement~\cite{waas_remote_2026}.

Addressing the remaining challenges in photon collection efficiency and emission rate requires photonic cavities that enhance light-matter coupling via the Purcell effect. While nanophotonic cavities enable high cooperativity through small mode volumes \cite{codreanu_above-unity_2026,yama_scalable_2026}, they face fabrication challenges and limited tunability. Open Fabry-Pérot microcavities \cite{janitz_cavity_2020} provide an alternative approach, combining tunability for spectral and spatial matching of inhomogeneous solid-state emitters with high finesse for strong light-matter coupling. Recent demonstrations have achieved Purcell enhancement of the zero phonon line of NV centers \cite{riedel_deterministic_2017,ruf_resonant_2021,yurgens_cavity-assisted_2024,fischer_spin-photon_2025}, group IV color centers~\cite{bayer_optical_2023,zifkin_lifetime_2024,berghaus_cavity-enhanced_2025}, and cooperative coupling of SnV centers in diamond \cite{herrmann_coherent_2023}.  

However, efficient operation in the diamond-like regime -- where the cavity field is concentrated inside the diamond membrane and Purcell enhancement is maximized -- has remained challenging due to scattering losses from surface roughness at the air-diamond interface \cite{dam_optimal_2018,flagan_diamond-confined_2022,korber_scanning_2023}. Previous work has demonstrated Purcell enhancement primarily in air-like cavity configurations, where the field antinode resides in the air gap, reducing sensitivity to surface quality but also limiting the achievable Purcell factor. 

Operating tunable microcavities at the mechanical stability and low temperatures required for group IV color centers poses further challenges, and earlier work remained largely vibration-limited to moderate finesse values while operating at temperatures above \SI{4}{\kelvin}, outside the regime required for long-lived electron spin coherence.

Here, we overcome these challenges and realize an open Fabry--Pérot microcavity platform with picometer-scale stability in a low-noise millikelvin cryostat and demonstrate cavity-enhanced light-matter coupling of SnV centers in a low-roughness diamond membrane. We realize diamond membranes with sub-nanometer surface roughness ($S_q = 0.2$~nm rms), enabling low-loss operation in the diamond-like regime. This yields an effective Purcell factor $C_0=4.1$, two-fold improved over air-like modes. We further demonstrate a coherent cooperativity $C=4.0$ with $96\%$ extinction contrast, and show spin-resolved cavity extinction with a spin contrast $C_\mathrm{spin}=91\%$ under an applied magnetic field. This marks a key step towards cavity-enhanced optical spin readout and spin--photon entanglement and establishes SnV centers coupled to open microcavities as a promising route towards scalable spin--photon interfaces.

%% file: Chapters/2_Setup.tex
\section{Experimental Setup}
\begin{figure}
    \centering
    \includegraphics[width=\columnwidth]{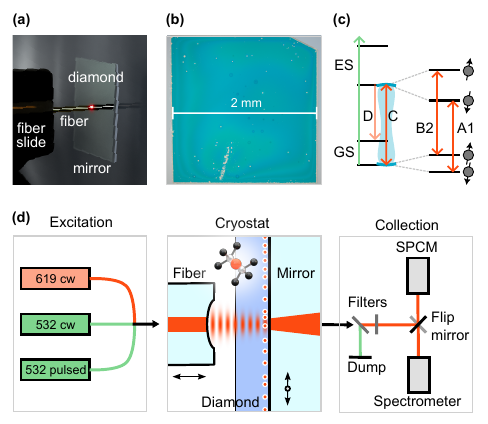}
    \caption{A cryogenic microcavity platform for solid-state quantum optics. (a) Photograph of the fiber-based plano-concave Fabry--P\'{e}rot cavity, showing the diamond membrane on the planar mirror and the fiber mirror mounted on a fiber slide. (b) Optical microscope image of the $2\times2$\,\si{\milli\meter\squared} diamond membrane bonded onto the planar cavity mirror. (c) Level scheme of the SnV center with the C~transition addressed by the cavity resonance. An applied magnetic field resolves the two spin-selective optical transitions. (d) Cavity-integrated SnV centers are excited off-resonantly or resonantly probed through the fiber mirror. Emitted fluorescence or transmitted probe light is spectrally filtered and detected using either a fiber-coupled spectrometer or a single-photon counting module (SPCM).}
    \label{fig:setup}
\end{figure}
Our experimental setup [Fig. 1(a)] combines a fiber-based Fabry-Pérot microcavity platform (Qlibri GmbH) integrated into a table-top closed-cycle dilution refrigerator (Qinu GmbH). The Fabry-Pérot cavity is formed between a $\mathrm{CO_2}$ laser-machined concave fiber mirror (radius of curvature \SI{30}{\micro\meter}) and a planar mirror onto which the diamond membrane is van-der-Waals bonded [Fig. 1(b)]. A coarse and a fine piezoelectric actuator acting on the fiber provide precise control over the cavity length and thus the cavity resonance frequency. The planar mirror and sample sit inside an $xy$-scanning unit of monolithic flexure design, driven by two piezoelectric actuators, while cryo-compatible stepper motors extend the positioning range along all axes to several millimeters.

\begin{figure*}[t]
    \centering
    \includegraphics[width=\textwidth]{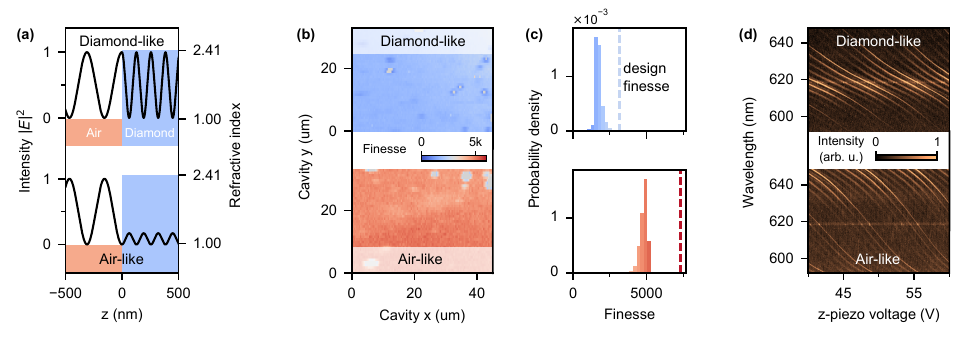}
    \caption{Hybrid air-diamond microcavity characterization. (a) Normalized intracavity electric-field intensity $|E|^2$ for a diamond-like mode (top) and an air-like mode (bottom), illustrating the contrasting field localizations across the air gap and diamond membrane. (b) Spatially resolved finesse maps acquired at lateral positions characteristic of the diamond-like and air-like modes. (c) Histograms of the finesse maps in (b), yielding median finesse values of $\mathcal{F}=1700$ and $\mathcal{F}=4840$ for the diamond-like and air-like modes, respectively; dashed lines indicate the corresponding design (lossless) values of $\mathcal{F}=3160$ and $\mathcal{F}=7320$. (d) Cavity dispersion scans recorded at the respective lateral positions, confirming the mode character: the diamond-like (air-like) mode exhibits a flat (steep) dispersion slope at $\lambda = \SI{619}{\nano\meter}$.}
    \label{fig:hybrid}
\end{figure*}

The diamond samples are prepared from electronic-grade single-crystal CVD diamond (Element Six Ltd.), polished to sub-nm surface roughness and implanted with $^{116}$Sn ions at a depth matched to the cavity field antinode, \SI{20}{\nano\meter} from the surface. After annealing, the plates are van-der-Waals bonded onto the planar cavity mirror and etched to a final thickness of \SIrange{0.8}{2.1}{\micro\meter}~\cite{heupel_fabrication_2020}. Two samples prepared by this procedure, differing in implantation dose and annealing conditions, are used throughout this work and referred to as sample~A and sample~B; full preparation details are given in Appendix~\ref{app:sample_prep}.

The fiber end facet and planar mirror are coated with identical custom distributed Bragg reflector, corresponding to a design finesse of $\mathcal{F} = 5040$ for the empty (air-filled) cavity; the finesse and achievable cooperativity actually realized in this hybrid air-diamond configuration are discussed in Sec.~\ref{sec:hybrid}. The microcavity platform is mounted on the millikelvin plate of the dilution refrigerator and integrated into the optical table. Without active feedback, the passive mechanical stability yields an rms cavity-length jitter of $\sigma_z = \SI{10.0}{\pico\meter}$ at the sample temperature $T_\mathrm{cav} = \SI{1.0}{\kelvin}$ -- sufficient for long-lived electron spin coherence -- and improves to \SI{4.5}{\pico\meter} at \SI{4}{\kelvin}. Both values lie well below the cavity's \SI{60}{\pico\meter} linewidth at this nominal finesse, leaving headroom to operate at even higher finesse. A full characterization is given in Appendix~\ref{app:stability}.

The cavity resonance is tuned to the C~transition of the SnV center, which connects the lower spin-orbit branches of the ground and excited-state manifolds [Fig.~\ref{fig:setup}(c)]. Applying an external magnetic field lifts the spin degeneracy of this transition, resolving two spin-selective optical transitions and thereby enabling a spin-photon interface. The cavity is excited and probed through the fiber mirror; transmitted or emitted light is spectrally filtered and detected by a fiber-coupled spectrometer or a single-photon counting module (SPCM) [Fig.~\ref{fig:setup}(d)].

%% file: Chapters/3_HybridCavity.tex
\section{Hybrid Diamond-Air Cavity}
\label{sec:hybrid}

Integrating a diamond membrane into the cavity modifies the effective mode volume $V_\mathrm{eff}$ and quality factor $Q$, and thus the Purcell factor
\begin{equation}
   F_P = \xi^2 \frac{3}{4\pi^2} \left(\frac{\lambda}{n}\right)^3 \frac{Q}{V_{\text{eff}}},
\end{equation}
in a non-trivial way. Here, $\xi \leq \cos(35^\circ)$ accounts for the polarization overlap between the cavity field and the emitter dipole, limited by the $\langle100\rangle$ crystal orientation of the diamond, which we incorporate into the Purcell factor. Depending on the local membrane thickness, the intracavity field interferes constructively or destructively inside the diamond, leading to diamond-like and air-like modes [Fig.~\ref{fig:hybrid}(a)], distinguished by whether a field antinode or node resides on the diamond-air interface. Referencing the mode volume to the emitter position, the Purcell factor becomes
\begin{equation}
    F_P = \frac{6}{\pi^3}\left(\frac{\lambda}{n_\mathrm{d}}\right)^2 \frac{\mathcal{F}}{w_0^2}\, \mathcal{I}_{A/D}^{-1},
    \label{eq:purcell}
\end{equation}
where $\mathcal{F}$ is the cavity finesse, $w_0$ the beam waist, and $\mathcal{I}_{A/D} = E^2_\mathrm{max,a}/(n_\mathrm{d} E^2_\mathrm{max,d})$ an intensity ratio quantifying the mode character, ranging between $1/n_\mathrm{d}$ (diamond-like) and $n_\mathrm{d}$ (air-like)~\cite{dam_optimal_2018}. Since $\mathcal{F}$ itself depends on the mode character through intensity-weighted mirror losses, diamond-like modes only outperform air-like ones if $n_\mathrm{d}^2\mathcal{F}_\mathrm{dia} > \mathcal{F}_\mathrm{air}$, a demanding requirement given that diamond-like modes feature a field antinode at the air-diamond interface, making them sensitive to scattering loss from surface roughness.

Although high finesse has been achieved in diamond-like configurations~\cite{hoy_jensen_cavity-enhanced_2020,flagan_diamond-confined_2022}, lifetime-shortening has so far been demonstrated only in air-like cavities~\cite{herrmann_coherent_2023,zifkin_lifetime_2024,yurgens_cavity-assisted_2024,fischer_spin-photon_2025}, as membrane roughness and the resulting scattering loss have limited efficient operation at diamond-like conditions despite their higher theoretical coupling strength. To overcome this limitation, we refine the polishing and etching procedure described in Appendix~\ref{app:sample_prep}, reducing the surface roughness from $S_q=\SI{2.8}{\nano\meter}$\,rms to $S_q=\SI{0.2}{\nano\meter}$\,rms. This enables reliable van-der-Waals bonding of the full $2\times2$~\si{\milli\meter\squared} membrane and low-loss operation in the diamond-like regime. 
A spatially resolved finesse map [Fig.~\ref{fig:hybrid}(b)] shows spatially homogeneous values across a $30\times45$\,\si{\micro\meter\squared} lateral area with no significant defects or scatterers, confirming the high membrane quality after processing. The histograms in Fig.~\ref{fig:hybrid}(c) yield median values of $\mathcal{F}_{\mathrm{dia}} = 1700 \pm 180$ and $\mathcal{F}_\mathrm{air} = 4840 \pm 250$, both reduced relative to the lossless design values of $\mathcal{F}=3160$ and $\mathcal{F}=7320$ (dashed lines), with the residual loss budget attributed primarily to fiber-mirror scattering and, for the diamond-like mode, additional scattering at the air-diamond interface (Appendix~\ref{app:losses}). These finesse values correspond to free-space outcoupling efficiencies of $\eta_\mathrm{fs} = 18(1)\,\%$ (air-like) and $37(4)\,\%$ (diamond-like), and we achieve an overall cavity-to-SPCM detection efficiency of up to $10\,\%$ (Appendix~\ref{app:losses}). We confirm the mode character at each lateral position independently via the cavity dispersion, measured from the broadband background fluorescence under off-resonant excitation with the attenuated \SI{619}{\nano\meter} laser line as a wavelength reference; diamond-like and air-like positions show characteristically flat and steep dispersion slopes, respectively [Fig.~\ref{fig:hybrid}(d)].

Using the measured finesse values and a mode waist of $w_0\approx\SI{1.2}{\micro\meter}$, Eq.~\eqref{eq:purcell} yields expected Purcell factors of $F_{P,\mathrm{dia}}\lesssim23$ and $F_{P,\mathrm{air}}\lesssim11$ -- a factor-of-two advantage for diamond-like modes despite their lower finesse. The accompanying flatter dispersion relaxes the cavity stability requirements, rendering it less susceptible to acoustic noise. The careful surface treatment of the membrane thus achieves $n_\mathrm{d}^2\,\mathcal{F}_{\mathrm{dia}} > \mathcal{F}_{\mathrm{air}}$, establishing diamond-like operation as the superior regime in our experiment. We verify this prediction experimentally in Sec.~\ref{sec:purcell} by directly comparing the Purcell factor and cooperativity achieved for both mode characters.

%% file: Chapters/4_Purcell.tex
\section{Lifetime shortening in a hybrid cavity}
\label{sec:purcell}

\begin{figure}[t]
    \centering
    \includegraphics[width=\columnwidth]{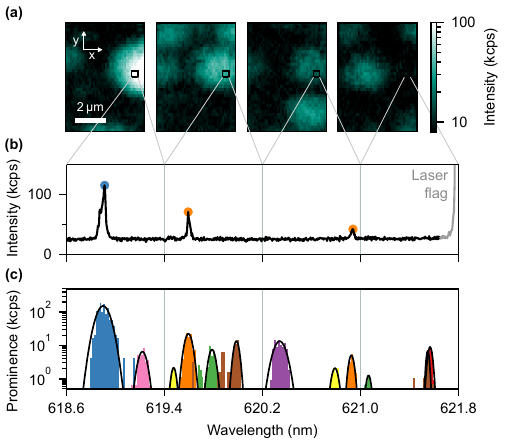}
    \caption{Cavity-enhanced hyperspectral characterization of SnV centers under off-resonant excitation. (a) Fluorescence map acquired over a $5 \times 7\,\SI{}{\micro\meter}^2$ lateral area; the color scale represents peak fluorescence intensity. (b) Representative spectrum at a single lateral position obtained by cavity scanning showing three spectrally isolated fluorescence peaks and a laser flag for wavelength reference. (c) Detected fluorescence signatures grouped by lateral position, wavelength, and peak prominence. Matching colors indicate features assigned to the C and D transitions of the same SnV center. 
    Peak widths reflect the statistical spread of peak positions across individual time traces, and thus the cavity linewidth, rather than the intrinsic linewidth of the transition itself.}
    \label{fig:hyperspectral}
\end{figure}

\begin{figure*}[t]
    \centering    \includegraphics[width=\textwidth]{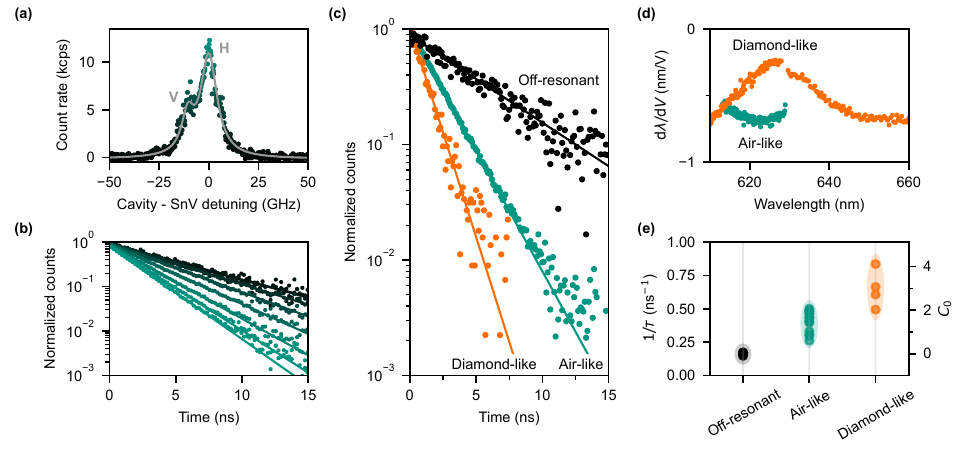}
    \caption{Purcell enhancement of SnV centers. (a) Cavity resonance scan over the brightest identified SnV center. The fluorescence as a function of cavity detuning reveals a double-peaked feature arising from the two orthogonal polarization eigenmodes (H and V) of the cavity, fitted with a double Lorentzian. The extracted linewidth of \SI{8.8\pm1.3}{\giga\hertz} is consistent with the cavity linewidth independently determined from resonant laser scans. (b) Excited-state lifetime measurements as a function of cavity-emitter detuning, with color encoding the detuning (green: on resonance, black: far off resonance). The on-resonance lifetime is Purcell-shortened to $\tau_c = \SI{2.07\pm0.01}{\nano\second}$ compared to the free-space lifetime $\tau_0 = \SI{5.90\pm0.11}{\nano\second}$, yielding an effective Purcell factor of $C_0 = \SI{1.85\pm0.05}{}$ at an air-like cavity position. (c) On-resonance lifetime measurement at a diamond-like cavity position, shown alongside the air-like data for comparison. The Purcell-shortened lifetime is further reduced to $\tau_c = \SI{1.19\pm0.02}{\nano\second}$, corresponding to $C_0 = \SI{4.1\pm0.1}{}$. (d) Dispersion curves of the fundamental cavity mode at both lateral positions. The slope magnitude exhibits a minimum (maximum) near \SI{619}{\nano\meter}, confirming the diamond-like (air-like) mode character. (e) Summary of all measured lifetimes with corresponding effective Purcell factors on the secondary axis, referenced to the mean free-space lifetime $\bar{\tau}_0 = \SI{6.1\pm0.4}{\nano\second}$. SnV centers at diamond-like positions consistently yield larger $C_0$.}
    \label{fig:purcell}
\end{figure*}

To spatially and spectrally isolate individual SnV centers, we exploit the lateral scanning capability of our microcavity platform together with sample A, whose lower implantation dose yields about one emitter on average within the cavity mode area. By scanning the cavity length across the expected C and D transition wavelengths while recording off-resonantly excited fluorescence on a lateral grid, we acquire hyperspectral fluorescence maps [Fig.~\ref{fig:hyperspectral}(a)], from which individual spectra such as the representative example in Fig.~\ref{fig:hyperspectral}(b) are extracted (Appendix~\ref{app:hyperspectral}). Grouping the resulting fluorescence peaks by lateral position and wavelength allows spatially overlapping but spectrally distinct features to be assigned to the C and D transitions of the same SnV center [Fig.~\ref{fig:hyperspectral}(c)]. For emitters where both transitions are resolved, the extracted ground-state splittings range from \SIrange{995}{1215}{\giga\hertz}, significantly exceeding the unstrained value of \SI{820}{\giga\hertz}~\cite{iwasaki_tin-vacancy_2017,karapatzakis_microwave_2024}. Combined with the broad spread of transition frequencies across the scan, this indicates that the implanted SnV centers reside in a defect-rich, strained crystal environment, likely due to implantation-induced damage and strain from the van der Waals bond.

We further study the identified emitters and characterize their cavity coupling by measuring their excited state lifetime using a pulsed green laser. An example scan of the cavity resonance over the brightest SnV center identified in Fig.~\ref{fig:hyperspectral} (blue) is shown in Fig.~\ref{fig:purcell}(a). The double-peaked feature arises from the polarization splitting of the cavity mode and indicates a suboptimal in-plane alignment between the cavity polarization and the C-transition dipole of this particular SnV center. To evaluate the effective Purcell factor $C_0 = \beta_\mathrm{tot} F_P = \tau_0/\tau_c - 1$, we measure both the off-resonant (free-space) lifetime $\tau_0$ and the on-resonance Purcell-shortened lifetime $\tau_c$. The total branching ratio $\beta_\mathrm{tot} = \beta_\mathrm{QE} \cdot \beta_\mathrm{C/D} \cdot \beta_\mathrm{DW} \approx 0.36$ combines the quantum efficiency $\beta_\mathrm{QE} \approx 0.8$~\cite{iwasaki_tin-vacancy_2017}, the C/D-transition branching ratio $\beta_\mathrm{C/D} \approx 0.8$~\cite{rugar_quantum_2021, lee_quantum_2025}, and the Debye--Waller factor $\beta_\mathrm{DW} \approx 0.57$~\cite{gorlitz_spectroscopic_2020}. The full spectral tunability allows us to measure the excited-state lifetime as a function of cavity-emitter detuning, as shown in Fig.~\ref{fig:purcell}(b), where the color encodes the detuning (green: on resonance, black: far off resonance). For this SnV center at an air-like cavity position, we obtain $\tau_0=\SI{5.90\pm0.11}{\nano\second}$ and$\tau_c=\SI{2.07\pm0.01}{\nano\second}$, yielding $C_0 = \tau_0/\tau_c - 1 = \SI{1.85\pm0.05}{}$.

To experimentally validate the enhanced cavity-emitter coupling in the diamond-like regime, we compare the Purcell-shortened lifetimes at air-like and diamond-like lateral positions [Fig.~\ref{fig:purcell}(c)]. We observe a nearly two-fold reduction to $\tau_c=\SI{1.19\pm 0.02}{\nano\second}$, corresponding to $C_0 = \SI{4.1\pm0.1}{}$. At both lateral positions, we acquire dispersion scans and track the resonant wavelength of the fundamental cavity mode as a function of cavity length [Fig.~\ref{fig:purcell}(d)]. The slope magnitude of each curve is minimal (maximal) near the SnV transition wavelength of \SI{619}{\nano\meter}, confirming the diamond-like (air-like) mode character at the respective emitter positions. All measured lifetimes and corresponding effective Purcell factors are summarized in Fig.~\ref{fig:purcell}(e), referenced to the mean free-space lifetime $\bar{\tau}_0=\SI{6.1\pm 0.4}{\nano\second}$. SnV centers at diamond-like positions consistently yield significantly larger $C_0$ despite the reduced finesse, as the stronger mode confinement overcompensates for additional mirror losses. This represents a key practical advantage: larger Purcell factors are achieved with lower finesse, relaxing the cavity length stability requirements and enabling more robust operation.

%% file: Chapters/5_Extinction.tex
\section{Resonant Extinction in Transmission}

\begin{figure*}[t]
    \centering
    \includegraphics[width=\textwidth]{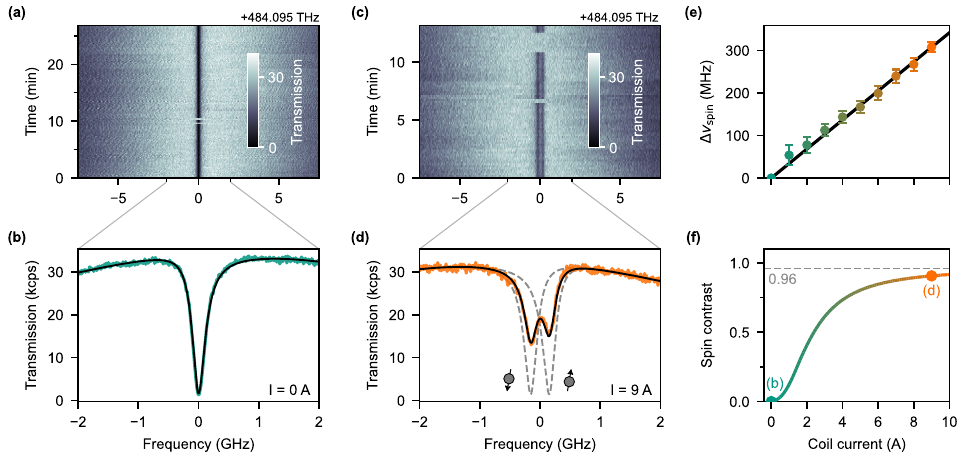}
    \caption{Cavity-enhanced extinction and Zeeman splitting of spin transitions in transmission.
    (a,c) Consecutive cavity transmission scans, recorded without and with applied magnetic field ($I_\mathrm{coil} = 0$ and $\SI{9}{\ampere}$), plotted as a function of laser frequency and time with transmission color-coded.
    Near the SnV C-transition at $f_\mathrm{SnV} \approx \SI{484.095}{\tera\hertz}$, a pronounced extinction feature appears at the cavity resonance with a 96\% contrast at zero field (a).
    Horizontal gaps mark intervals in which the emitter occupied a dark charge state and a repump pulse was applied.
    At maximum coil current, the extinction feature splits into two branches corresponding to the two spin-conserving transitions.
    (b,d) Corresponding averaged spectra: for each scan, the emitter frequency is determined and the spectra are aligned to emitter-centered detuning before averaging.
    At zero field (b), the data are described by a single-emitter cavity--QED fit (solid line).
    At $I_\mathrm{coil} = \SI{9}{\ampere}$ (d), a split cavity--QED fit (solid line) yields $\Delta\nu_\mathrm{spin} = \SI{308(12)}{\mega\hertz}$, with the individual spin-resolved contributions shown as dashed curves.
    (e) Extracted Zeeman splitting $\Delta\nu_\mathrm{spin}(I_\mathrm{coil})$ as a function of coil current, following the expected linear dependence (solid line) with slope $\Delta\nu_\mathrm{spin}/I_\mathrm{coil} = \SI{34.2(8)}{\mega\hertz\per\ampere}$.
    (f) Maximum achievable spin contrast $\mathcal{C}_\mathrm{spin}$ as a function of coil current for the present linewidth $\gamma_\mathrm{tot}/2\pi = \SI{52}{\mega\hertz}$, asymptotically approaching the cooperativity-limited contrast $\mathcal{C} = 0.96$ (dashed line).
    At the experimentally achieved splitting of $\Delta\nu_\mathrm{spin} = \SI{308}{\mega\hertz}$, the measured contrast is $\mathcal{C}_\mathrm{spin} = 0.91$.
    Markers indicate the operating points of panels (b) and (d).}
    \label{fig:extinction}
\end{figure*}

We next investigate cavity-enhanced extinction through resonant probing of the cavity-SnV system. To facilitate the detection of emitters, we use sample B, which exhibits higher spatial and spectral emitter density, and restrict the search to lateral positions near the diamond-like regime, where coupling is strongest. 
We sweep a weak, tunable probe laser across the cavity resonance and monitor the transmitted intensity, searching for extinction dips that indicate coherent cavity--emitter coupling -- an approach motivated by the high effective Purcell factors established from the lifetime measurements above, which predict a pronounced modification of the cavity transmission on resonance with an SnV center (Appendix~\ref{app:emitter_search}). The following measurements are performed on a single SnV center selected for its largest extinction contrast and highest charge stability under resonant illumination. All measurements in this and the following section were taken at $T=\SI{5.0}{\kelvin}$ rather than at the base temperature of $T=\SI{1.0}{\kelvin}$: the elevated temperature leads to a shortened spin relaxation time $T_1$ and thus suppresses optical spin-pumping, which otherwise distorts the extinction lineshape under an applied magnetic field. For consistency, we used the same elevated temperature also for the zero-field measurement.

Figure~\ref{fig:extinction}(a) shows consecutive cavity transmission scans near the SnV C-transition at $f_\mathrm{SnV} \approx \SI{484.095}{\tera\hertz}$. On resonance, the transition produces a pronounced extinction dip with \SI{96}{\percent} contrast, characteristic of the high-cooperativity regime. The probe power is chosen to be about $0.01\%$ of the saturation of the system, corresponding to a count rate of 30~kcps on the detector. Notably, no spectral diffusion is observable under these conditions, and the emitter shows only occasional charge state jumps on a 20~min timescale that are visible as missing extinction signals. The emitter can then be repumped by brief excitation with 532~nm light. Higher probing powers induce more frequent charge state jumps but do not compromise spectral stability.
Individual transmission traces are frequency-aligned to the emitter resonance, averaged, and fitted to a cavity-QED model (Appendix~\ref{app:cqed_fit}). Figure~\ref{fig:extinction}(b) shows the resulting averaged spectrum together with the best-fit model curve. From $N = 156$ valid scans, we obtain
\begin{equation}
    \{g, \kappa, \gamma_\mathrm{tot}\}/2\pi = \{0.84(2),\, 13.5(3),\, 0.052(16)\}\,\text{GHz},
\end{equation}
corresponding to $C = 4.0(13)$ and a resonant extinction contrast $\mathcal{C} = 0.96(2)$. The extracted linewidth is twice the transform-limited value of $\gamma_0/2\pi \approx 26$\,MHz, which we attribute to fast charge-noise-induced spectral diffusion associated with the shallow implantation depth. Extrapolating to $\gamma_\mathrm{tot} \to \gamma_0$ yields an ideal cooperativity of $C_0 = 8.0(5)$ and contrast of $\mathcal{C}_0 = 0.99(1)$.

%% file: Chapters/6_Zeeman.tex
\section{Optical Zeeman Splitting in Extinction}

A crucial step towards a spin--photon interface is magnetically lifting the degeneracy between the $m_s = \pm 1/2$ spin sublevels. This splitts the single extinction feature into two spin-selective transitions, enabling optical spin readout and spin-dependent transmission or reflection of single photons from the cavity. We apply a variable magnetic field using a superconducting coil centered on the optical axis, providing a fixed field orientation relative to the emitters. The coil current is varied over $I_\mathrm{coil} = 0$--\SI{9}{\ampere}, limited by superconductor quenching.

At the maximum coil current of $I_\mathrm{coil} = 9$\,A, the transmission traces [Fig.~\ref{fig:extinction}(c)] reveal two clearly resolved extinction branches corresponding to the two spin states, each with reduced contrast compared to the zero-field case. Averaging the emitter-centered spectra recorded at this current and fitting the result with a cavity--QED response for two split resonances [Fig.~\ref{fig:extinction}(d), Appendix~\ref{app:spin_contrast}] yields $\Delta\nu_\mathrm{spin}(I_\mathrm{coil} = 9\,\mathrm{A}) = 308(12)$\,MHz.
Repeating this analysis for each coil current yields the splitting $\Delta\nu_\mathrm{spin}(I_\mathrm{coil})$ shown in Fig.~\ref{fig:extinction}(e), which follows the expected linear dependence on coil current with a fitted slope of $\Delta\nu_\mathrm{spin}/I_\mathrm{coil} = 34.2(8)$\,MHz\,A$^{-1}$. The measured splitting is compatible with the expected optical Zeeman splitting for a moderately strained SnV center at the estimated magnetic field at the sample position (Appendix~\ref{app:expected_zeeman}).

The demonstrated spin-selective extinction enables spin--photon entanglement, with a fidelity determined by the spin contrast $\mathcal{C}_\mathrm{spin} = |T_{\uparrow} - T_{\downarrow}|/T_{g=0}$. In the limit of large Zeeman splitting, the two spin transitions are spectrally resolved and the contrast approaches the cooperativity-limited value $\mathcal{C} = 1 - 1/(1+C)^2$. At finite splitting, both transitions contribute at the probe frequency, reducing the achievable contrast; this reduction can be partially mitigated by probing at a finite detuning from the cavity resonance, where the asymmetric, Fano-like line shapes of the two transitions maximize the difference in transmission between the spin states. Maximizing $\mathcal{C}_\mathrm{spin}$ over this detuning for each coil current (Appendix~\ref{app:spin_contrast}) yields the optimum shown in Fig.~\ref{fig:extinction}(f) for the present linewidth $\gamma_\mathrm{tot}/2\pi = 52$\,MHz, which asymptotically approaches the cooperativity-limited contrast $\mathcal{C} = 0.96$ as the splitting increases.

At the experimentally achieved splitting of $\Delta\nu_\mathrm{spin} = 308$\,MHz, the optimum occurs at an emitter--cavity detuning of $\omega_{a,\downarrow(\uparrow)} - \omega_c \approx \mp 2.3$\,GHz, yielding $\mathcal{C}_\mathrm{spin} = 0.91$; only in the limit of infinite splitting does the optimal detuning approach zero, where probing on resonance with a single, spectrally isolated spin transition directly recovers the cooperativity-limited contrast. These results establish spectrally resolved, spin-selective extinction as the basis for optical spin readout. 

%% file: Chapters/7_Conclusion.tex
\section{Conclusion and Outlook}

This work establishes a fiber-based Fabry–Pérot microcavity platform as a promising architecture for quantum network nodes based on group-IV color centers in diamond. By integrating diamond-membranes with surface roughness below 0.2 nm rms, we access the diamond-like regime of cavity operation, achieving cooperativities exceeding $C=4$, despite reduced finesse--a decisive practical advantage that relaxes cavity length stability requirements while enhancing emitter–cavity coupling. Resolving spin-selective transitions with $91\%$ contrast provides the foundation for cavity-mediated single-shot spin readout and spin–photon entanglement generation. 
Notably, the high efficiency of the cavity-enhanced spin-photon interaction enables high-fidelity spin state readout even under off-axis magnetic fields. Such field orientations are highly beneficial for accelerated microwave spin control of SnV centers \cite{pieplow_efficient_2024}, but lead to low optical cyclicity \cite{rosenthal_single-shot_2024}. In our experiments, overall photon detection efficiencies of up to $10\%$ were achieved (including cavity outcoupling, cryostat and setup 
transmission, and detector quantum efficiency), such that high-fidelity 
single-shot readout appears feasible for optical cyclicities $> 10$, rendering nearly any magnetic field orientation compatible with deterministic readout. Open-microcavity architectures have been shown to support considerably higher collection efficiencies~\cite{tomm_bright_2021,ding_high-efficiency_2025}, suggesting that further gains are achievable through optimized optics, as well as optimized cavity directionality.

Together, these results demonstrate that open microcavities can achieve the strong cooperativity, spin resolution, and operational flexibility required for practical quantum network nodes. This positions SnV centers in open cavities as a compelling alternative to nanophotonic approaches, with a clear pathway toward high-fidelity spin-photon entanglement for quantum networks.


%% file: Chapters/8_Appendix.tex
\section{Diamond sample preparation}
\label{app:sample_prep}
\begin{figure}
    \centering
    \includegraphics[width=\columnwidth]{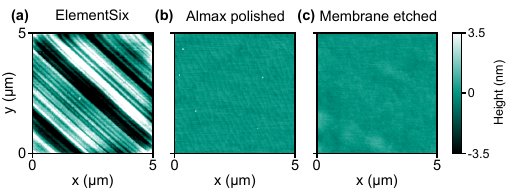}
    \caption{Surface roughness characterization by atomic force microscopy. (a) As-received Element Six diamond substrate with $S_q = \SI{2.8}{\nano\meter}$ rms, showing pronounced polishing scratches. (b) After Almax polishing, the surface roughness is reduced to $S_q = \SI{0.2}{\nano\meter}$ rms. (c) After etching to a \SIrange{1}{2}{\micro\meter} thin membrane, the sub-nm roughness is preserved at $S_q = \SI{0.2}{\nano\meter}$ rms. All maps are acquired over a $5\times5\,\si{\micro\meter\squared}$ area.}
    \label{fig:AFM}
\end{figure}
The diamond membranes are produced from $2\times2\times0.04$~\si{\milli\meter\cubed} electronic-grade single-crystal CVD diamond plates supplied by Element Six Ltd. and polished to sub-nm surface roughness by Almax EasyLab. Figure~\ref{fig:AFM} shows AFM height maps at each stage of the preparation process. The as-received substrate exhibits pronounced polishing scratches with $S_q = \SI{2.8}{\nano\meter}$ rms [Fig.~\ref{fig:AFM}(a)], which are eliminated after Almax polishing, reducing the roughness to $S_q = \SI{0.2}{\nano\meter}$ rms [Fig.~\ref{fig:AFM}(b)]. All full-sized plates were successfully van-der-Waals bonded onto the planar cavity mirror prior to etching. Sample~A (B) was implanted with $^{116}$Sn ions at a kinetic energy of \SI{75}{\kilo\eV} (\SI{30}{\kilo\eV}) and a dose of \SI{1e9}{\per\centi\meter\squared} (\SI{4e9}{\per\centi\meter\squared}). The \SI{75}{\kilo\eV} energy places the ions at a projected range of \SI{24}{\nano\meter}, closely matching the cavity field antinode located \SI{20}{\nano\meter} from the diamond surface. For sample~B, the lower energy of \SI{30}{\kilo\eV} was applied at normal incidence to exploit ion channeling along the crystal axis, reducing nuclear stopping and producing a less damaged crystal environment, while maintaining a comparable implantation depth~\cite{bushmakin_two-photon_2025}. Subsequently, sample~A (B) was annealed at \SI{1200}{\celsius} (\SI{1450}{\celsius}) for 4\,hrs (2\,hrs) to promote color center formation and healing of implantation-induced lattice damage. After tri-acid and Piranha cleaning, the full $2\times2$~\si{\milli\meter\squared} plates are van-der-Waals bonded onto the planar cavity mirror and etched down to \SIrange{0.8}{2.1}{\micro\meter} by inductively coupled plasma reactive ion etching (ICP-RIE), following the procedure described in \cite{heupel_fabrication_2020}. The sub-nm surface roughness is preserved after etching, with the membrane surface exhibiting $S_q = \SIrange{0.2}{0.3}{\nano\meter}$ rms over a $5\times5$~\si{\micro\meter\squared} area [Fig.~\ref{fig:AFM}(c)].

\section{Cavity platform, mechanical stability and thermalization}
\label{app:stability}
The microcavity platform (Qlibri) is integrated into a table-top closed-cycle dilution refrigerator (Qinu), whose gas handling -- including the liquefier, vacuum pumps, and compressors -- is located in a separate room. A coarse and a fine piezoelectric actuator acting on the fiber provide control over the cavity length and thus the cavity resonance frequency, while the planar mirror and sample sit inside an $xy$-scanning unit of monolithic flexure design, driven by two piezoelectric actuators, with cryo-compatible stepper motors extending the positioning range along all axes to several millimeters. Together with the passive mechanical damping of this rigid, flexure-based assembly, the remote gas handling provides the low-vibration cryogenic environment required for stable, high-finesse cavity operation without active feedback.

\begin{figure}
    \centering
    \includegraphics[width=\columnwidth]{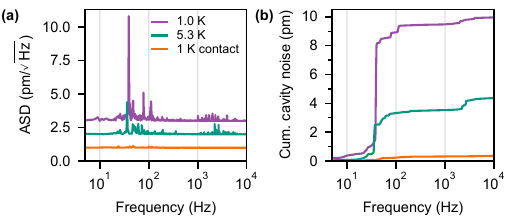}
    \caption{Passive mechanical stability of the cavity. (a) Amplitude spectral density (ASD) 
    of the cavity length noise and (b) cumulative cavity length noise as a function of frequency, 
    measured at $T_\mathrm{cav} = \SI{1.0}{\kelvin}$ (purple), $T_\mathrm{cav} = \SI{5.3}{\kelvin}$ 
    (teal), and at \SI{1.0}{\kelvin} in contact mode (orange). Without active feedback, the 
    integrated length jitter amounts to $\sigma_z = \SI{10.0}{\pico\meter}$ rms at \SI{1.0}{\kelvin} 
    and $\sigma_z = \SI{4.5}{\pico\meter}$ rms at \SI{5.3}{\kelvin}, dominated by low-frequency 
    mechanical noise below \SI{100}{\hertz}. In contact mode, the jitter is suppressed to 
    $\sigma_z = \SI{0.6}{\pico\meter}$ rms.}
    \label{fig:stability}
\end{figure}

The mechanical stability of the cavity is characterized by locking the cavity resonance to a narrow-linewidth laser via a side-of-fringe feedback loop operated at the lowest usable integral gain, providing slow drift compensation while leaving the broadband noise spectrum unaffected. The near-equivalence of this drift-compensated and the true passive stability was verified independently by comparing their amplitude spectral densities and integrated rms values, which were found to agree to within \SI{10}{\percent}. The cavity-length noise is extracted from the transmitted intensity time trace recorded over \SI{1}{\second} by converting the signal to a length displacement using the laser wavelength $\lambda$, the cavity finesse $\mathcal{F}$, and the peak transmission amplitude. Spectral analysis via a fast Fourier transform yields the amplitude spectral density (ASD) shown in Fig.~\ref{fig:stability}(a), and cumulative integration over frequency yields the integrated rms cavity-length noise shown in Fig.~\ref{fig:stability}(b).

Measurements were performed at two operating points of the dilution refrigerator. In the 4\,K mode, the cavity temperature stabilizes at $T_\mathrm{cav} = \SI{5.3}{\kelvin}$, yielding a passive rms length jitter of $\sigma_z = \SI{4.5}{\pico\meter}$. Operating the cryostat in its millikelvin (condensation) mode with an additional copper thermal link to the mixing-chamber plate reduces the cavity temperature to $T_\mathrm{cav} = \SI{1.0}{\kelvin}$. However, the additional mechanical connection to the cold plate increases the rms jitter to $\sigma_z = \SI{10.0}{\pico\meter}$, with the excess noise dominated by a spectral component near \SI{39}{\hertz} associated with the pulse-tube cryocooler. In both cases, the noise is concentrated below \SI{100}{\hertz}, as visible in the ASD and the cumulative noise plots. When the fiber mirror and the planar sample mirror are brought into mechanical contact, the resulting monolithic configuration substantially enhances the rigidity of the assembly and suppresses differential mirror motion. In this contact mode at $T_\mathrm{cav} = \SI{1.0}{\kelvin}$, the rms cavity-length jitter is reduced to $\sigma_z = \SI{0.6}{\pico\meter}$ ($\sigma_{z,\mathrm{cum}} = \SI{0.4}{\pico\meter}$ integrated up to \SI{10}{\kilo\hertz}) at a finesse of $\mathcal{F} = 5200$, corresponding to a length jitter of approximately \SI{1}{\percent} of the cavity linewidth.

We note that the base temperature reached in either mode varies between cooldowns; $T_\mathrm{cav}=\SI{1.0}{\kelvin}$, at which the stability data below were recorded, is the lowest cavity temperature achieved to date in the condensation mode, whereas typical cooldowns in this mode thermalize the cavity to $T_\mathrm{cav}\approx 1$--\SI{2}{\kelvin}.

\section{Hybrid cavity loss and detection efficiency}
\label{app:losses}

The round-trip loss $\mathcal{L} = 2\pi/\mathcal{F}$ is decomposed into 
mirror contributions and a residual term. The fiber mirror contributes 
$\mathcal{L}_\mathrm{M,a} = \SI{623}{ppm}$ to the round-trip loss. The planar 
mirror loss is weighted by the local field intensity at the diamond--air 
interface: in the air-like limit, $\mathcal{L}_\mathrm{M,d}/n_\mathrm{d} = 
\SI{235}{ppm}$, while in the diamond-like limit, $\mathcal{L}_\mathrm{M,d} 
\cdot n_\mathrm{d} = \SI{1364}{ppm}$, where $\mathcal{L}_\mathrm{M,d} = 
\SI{566}{ppm}$ and $n_\mathrm{d} = 2.41$. These yield design finesses of 
$\mathcal{F} = 7320$ and $\mathcal{F} = 3160$ for the air-like and diamond-like 
limits, respectively. The measured median finesses fall below these values, 
with residual losses of $\mathcal{L}_\mathrm{res} \approx \SI{442}{ppm}$ 
(air-like) and $\approx \SI{1717}{ppm}$ (diamond-like) beyond the two-mirror 
budget. Since the air-like residual is mostly insensitive to the diamond surface, it is most plausibly attributed to scattering at the fiber mirror. Assuming this 
contribution carries over unchanged to the diamond-like configuration, the 
excess residual of $\approx \SI{1275}{ppm}$ is attributed to additional 
scattering losses at the air-diamond interface, where the field antinode of the 
diamond-like mode enhances sensitivity to surface roughness. A complete loss 
attribution would require finesse measurements as a continuous function of mode 
character, accessible, for example, by varying the laser wavelength~\cite{hessenauer_cavity_2025} or the local diamond 
thickness~\cite{korber_scanning_2023,herrmann_laser-cut_2025}.

Since the planar-mirror transmission is the only loss channel that couples 
into the observed free-space port, the outcoupling efficiency is simply its 
fraction of the total measured round-trip loss, $\eta_\mathrm{fs} = 
\mathcal{L}_\mathrm{M,d}^{(\cdot)}/\mathcal{L}$. With $\mathcal{L} = 
2\pi/\mathcal{F}_\mathrm{meas} \approx \SI{1300}{ppm}$ (air-like) and 
$\approx\SI{3700}{ppm}$ (diamond-like), this yields $\eta_\mathrm{fs}\approx
\SI{18}{\percent}$ and $\approx\SI{37}{\percent}$, respectively. The 
subsequent collection efficiency onto the SPCM -- set by detector quantum 
efficiency, aperture, and free-space optics loss/residual reflections -- is 
$\eta_\mathrm{det}\approx\SI{26}{\percent}$, so the overall efficiency 
$\eta_\mathrm{fs}\eta_\mathrm{det}$ drops to $\approx\SI{5}{\percent}$ 
(air-like) and $\approx\SI{10}{\percent}$ (diamond-like).

\section{Hyperspectral imaging and spectral calibration}
\label{app:hyperspectral}
The hyperspectral fluorescence maps of Fig.~\ref{fig:hyperspectral} are acquired by exciting the sample through the fiber mirror with a continuous-wave green laser at \SI{2}{\milli\watt} and collecting the cavity-coupled fluorescence via the free-space detection path onto a single-photon counting module. Photon count-rate traces are recorded on a lateral grid while the cavity length is scanned across the expected SnV C and D transition wavelengths, spanning \SIrange{618.6}{621.8}{\nano\meter} with an integration time of \SI{7}{\second} per pixel; the scan is restricted to lateral positions where the cavity remains predominantly air-like, with a finesse exceeding 4000. The resulting time traces are converted into approximate spectra through a two-point wavelength calibration: a laser flag at a known wavelength of \SI{621.8}{\nano\meter} provides the first reference, while the fluorescence wavelength of an identified SnV spectral feature serves as the second. A two-dimensional peak-finding algorithm is applied to detect all fluorescence signatures across the full map and extract their lateral position and wavelength; the resulting features are grouped by wavelength and peak prominence to identify spatially overlapping but spectrally distinct transitions belonging to the same emitter.

\section{Off-resonant saturation measurement}
\label{app:saturation}
\begin{figure}
    \centering
    \includegraphics[width=\columnwidth]{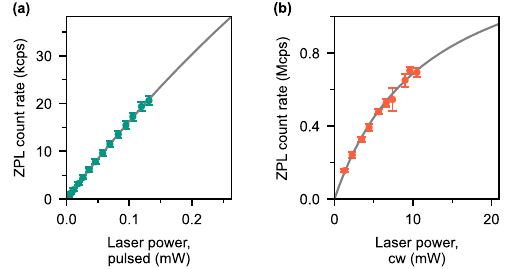}
    \caption{Saturation of the ZPL count rate for the brightest emitter identified in the hyperspectral imaging of Fig.~\ref{fig:hyperspectral}, under off-resonant excitation at \SI{532}{\nano\meter} with the cavity tuned for maximum spatial overlap. (a)~Pulsed excitation (repetition rate $f_\mathrm{rep} = \SI{80}{\mega\hertz}$). (b)~Continuous-wave excitation. Markers show the measured ZPL count rate $I$ vs.\ excitation power $P$ (mean $\pm$ standard deviation); gray lines are fits to the saturation model of Eq.~\eqref{eq:saturation}, yielding $I_\mathrm{s}^\mathrm{pulsed} = 222(18)$~kcts/s and $P_\mathrm{s}^\mathrm{pulsed} = 1.27(11)$~mW for pulsed excitation, and $I_\mathrm{s}^\mathrm{cw} = 1.50(11)$~Mcts/s and $P_\mathrm{s}^\mathrm{cw} = 11.8(13)$~mW for cw excitation.}
    \label{fig:saturation}
\end{figure}

We study the saturation behavior of the brightest emitter identified in the hyperspectral imaging of Fig.~\ref{fig:hyperspectral} by spatially tuning the cavity for maximum overlap and recording the ZPL count rate as a function of excitation power under pulsed and cw off-resonant excitation at \SI{532}{\nano\meter} (Fig.~\ref{fig:saturation}). Both data sets follow the standard saturation model
\begin{equation}
    I(P) = I_\mathrm{s}\,\frac{P/P_\mathrm{s}}{1+P/P_\mathrm{s}},
    \label{eq:saturation}
\end{equation}
where $I_\mathrm{s}$ is the saturation count rate and $P_\mathrm{s}$ the saturation power, defined via $I(P_\mathrm{s}) = I_\mathrm{s}/2$. For pulsed excitation [Fig.~\ref{fig:saturation}(a)], we obtain $I_\mathrm{s}^\mathrm{pulsed} = 222(18)$~kcts/s and $P_\mathrm{s}^\mathrm{pulsed} = 1.27(11)$~mW; for cw excitation [Fig.~\ref{fig:saturation}(b)], the fit yields $I_\mathrm{s}^\mathrm{cw} = 1.50(11)$~Mcts/s and $P_\mathrm{s}^\mathrm{cw} = 11.8(13)$~mW.

The different saturation values directly reflect the two excitation schemes. Under cw excitation the emitter is repumped immediately after each emission event, so $I_\mathrm{s}^\mathrm{cw} = \eta_\mathrm{det}/\tau_c$ is set by the (Purcell-shortened) excited-state lifetime $\tau_c$. Under pulsed excitation, with a pulse duration much shorter than $\tau_c$, at most one excitation can occur per pulse, so $I_\mathrm{s}^\mathrm{pulsed} = \eta_\mathrm{det} f_\mathrm{rep}$ is instead set by the repetition rate. The expected ratio $I_\mathrm{s}^\mathrm{pulsed}/I_\mathrm{s}^\mathrm{cw} = f_\mathrm{rep}\tau_c \approx 0.17$, for $\tau_c \approx \SI{2.1}{\nano\second}$, agrees well with the measured ratio of $0.15(2)$; the same reasoning accounts for the measured saturation-power ratio $P_\mathrm{s}^\mathrm{pulsed}/P_\mathrm{s}^\mathrm{cw} = 0.11(2)$ (expected $\approx 0.17$), with the larger deviation here attributed to additional uncertainty in the absolute power calibration. Comparing $I_\mathrm{s}^\mathrm{pulsed}$ to the maximum possible detected rate $f_\mathrm{rep}$ yields an overall emitter-to-detector efficiency of $\eta_\mathrm{det} \approx 0.3\%$, below the value expected from the setup transmission and measured cooperativity for an air-like cavity mode, which we attribute to charge-state dynamics reducing the average bright-state population under off-resonant excitation.

\section{Resonant extinction search and acquisition procedure}
\label{app:emitter_search}
Resonant emitter search is performed by driving the cavity with a weak coherent field of 1--10\,pW and recording the transmitted intensity on an SPCM while sweeping the laser frequency across the cavity resonance near the SnV C-transition at $f^0_\mathrm{SnV} = 484.134$\,THz~\cite{narita_multiple_2023}. SnV centers are identified by pronounced extinction dips in the transmission profile. Each sweep spans 25\,GHz, and the cavity resonance is stepped across the scan range over consecutive sweeps to ensure full spectral coverage given the finite cavity linewidth of 8--16\,GHz. Where significant polarization splitting is present, both polarization modes are probed independently. Prior to each sweep, a 532\,nm repump pulse (10\,ms, 10\,mW) reinitializes emitters from dark charge states. The search is conducted at multiple lateral positions within the full lateral $10 \times 10$\SI{}{\micro\meter\squared} piezo scan window, with the spectral scan range shifted and the sample coarsely repositioned via stepper motors if no emitter is detected within a 100--200\,GHz band.

Once a target emitter is selected, transmission spectra for the extinction measurement are recorded under the same weak-probe conditions. The symmetric mirror coatings yield strong on-resonance transmission, permitting count rates above 100\,kcps while remaining well below emitter saturation; we operate at approximately 30\,kcps to preserve emitter charge stability over multiple consecutive scans.

\section{Cavity-QED fit model and analysis}
\label{app:cqed_fit}

Each transmission trace exhibiting an extinction feature is fitted with a model that accounts for the cavity-QED response of the coupled emitter--cavity system and a constant detector background. The full fit function for the detected intensity is
\begin{equation}
    I(\omega_l; \Theta) = I_\mathrm{cQED}(\omega_l; \Theta) + I_\mathrm{bg,dc},
    \label{eq:cqed_fit}
\end{equation}
where $\Theta = \{A, g, \kappa, \gamma_\mathrm{tot}, \omega_c, \omega_a\}$ collects all free parameters of the fit. The cavity-QED contribution is
\begin{equation}
    I_\mathrm{cQED}(\omega_l) = A \left| \frac{(\kappa/2)^2}{\frac{\kappa}{2} - i(\omega_l - \omega_c) + \frac{g^2}{\frac{\gamma_\mathrm{tot}}{2} - i(\omega_l - \omega_a)}} \right|^2,
    \label{eq:icqed}
\end{equation}
where $A$ is the on-resonance transmission amplitude, $g$ is the emitter--cavity coupling rate, $\kappa$ is the cavity intensity decay rate, $\gamma_\mathrm{tot} = \gamma_0 + 2\gamma^\star$ is the total emitter linewidth including pure dephasing, and $\omega_c$ and $\omega_a$ are the cavity and emitter resonance frequencies, respectively. The constant offset $I_\mathrm{bg,dc} = 500$\,cts/s accounts for detector dark counts (200\,cts/s intrinsic) and residual ambient illumination (300\,cts/s).

For the zero-field cooperativity analysis, each transmission trace is first fitted individually to determine the emitter frequency $\omega_a$ from the extinction feature. The spectrum is then mapped onto the emitter-centered detuning $\omega_l - \omega_a$ and interpolated onto a common frequency grid. The aligned traces are averaged to obtain a single spectrum with improved signal-to-noise ratio, which is fitted using Eqs.~\eqref{eq:cqed_fit} and~\eqref{eq:icqed}. The cQED parameters quoted in the main text are the best-fit values obtained from this averaged spectrum. Their uncertainties are estimated as the standard deviation of the corresponding parameter across the $N$ individual scans entering the average, thereby reflecting scan-to-scan variability from spectral diffusion and other experimental fluctuations rather than the formal fit covariance of the averaged trace.

\section{Split cavity-QED model and spin-contrast optimization}
\label{app:spin_contrast}

\begin{figure}[t]
    \centering
    \includegraphics[width=\columnwidth]{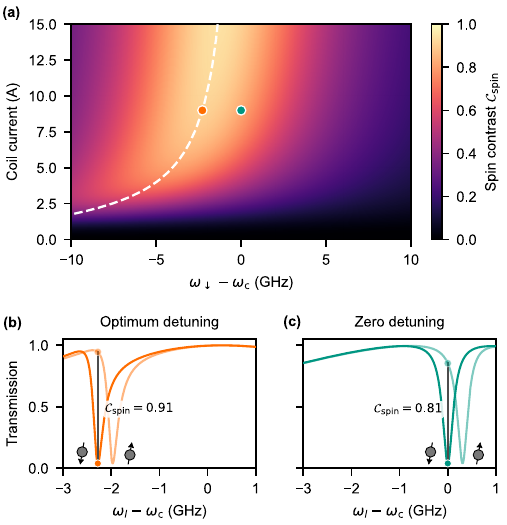}
    \caption{Spin-contrast optimization. (a) Theoretical spin contrast $\mathcal{C}_\mathrm{spin}$ as a function of coil current and cavity detuning $\delta = \omega_{a,\downarrow} - \omega_c$ from the probed, lower-energy $\downarrow$ transition, for the coupling parameters quoted in the main text. The dashed line traces the optimal detuning $\delta_\mathrm{opt}(I_\mathrm{coil})$ maximizing $\mathcal{C}_\mathrm{spin}$, approaching $\delta=0$ at large splitting. Markers indicate the two $I_\mathrm{coil}=9$\,A operating points examined in panels (b,c). (b,c) Transmission spectra vs.\ $\omega_l-\omega_c$ for the emitter in the probed $\downarrow$ state and the unprobed $\uparrow$ state, at the optimal detuning (b) and at zero detuning (c).}
    \label{fig:spin_contrast_map}
\end{figure}

The transmission spectra at finite coil current [Fig.~\ref{fig:extinction}(d)] are fitted with a split extension of the model in Appendix~\ref{app:cqed_fit}: the cavity-QED term of Eq.~\eqref{eq:icqed} is replaced by an equal-weight sum of two such terms, sharing $g$, $\kappa$, and $\gamma_\mathrm{tot}$ but with atomic frequencies $\omega_{a,\downarrow}$ and $\omega_{a,\uparrow} = \omega_{a,\downarrow} + 2\pi\Delta\nu_\mathrm{spin}$,
\begin{equation}
\begin{split}
    I_\mathrm{split}(\omega_l) = \tfrac{1}{2}\big[I_\mathrm{cQED}(\omega_l;\omega_c,\omega_{a,\downarrow}) \\
    {}+ I_\mathrm{cQED}(\omega_l;\omega_c,\omega_{a,\uparrow})\big] + I_\mathrm{bg,dc},
\end{split}
\label{eq:split_cqed}
\end{equation}
reflecting the equal population of both, unpolarized spin states. Fitting Eq.~\eqref{eq:split_cqed} at each coil current gives the splitting $\Delta\nu_\mathrm{spin}(I_\mathrm{coil})$ shown in Fig.~\ref{fig:extinction}(e).

For the contrast analysis of Fig.~\ref{fig:extinction}(f), the probe is instead fixed on resonance with the $\downarrow$ transition, $\omega_l = \omega_{a,\downarrow}$, and the cavity detuning $\delta = \omega_{a,\downarrow} - \omega_c$ is scanned. $T_\downarrow(\delta)$ and $T_\uparrow(\delta)$ then follow from Eq.~\eqref{eq:icqed} at atomic frequency $\omega_{a,\downarrow}$ and $\omega_{a,\uparrow}$, respectively [Fig.~\ref{fig:spin_contrast_map}(b,c)], with $\mathcal{C}_\mathrm{spin}(\delta) = |T_\uparrow(\delta) - T_\downarrow(\delta)|$. On resonance, both transitions contribute at the probe frequency, limiting the contrast; detuning the cavity leaves the addressed $\downarrow$ response nearly unchanged while pushing the already off-resonant $\uparrow$ branch further onto the wing of its asymmetric, Fano-like line shape, improving $\mathcal{C}_\mathrm{spin}$. For each coil current, $\mathcal{C}_\mathrm{spin}(\delta)$ is maximized over $\delta$, giving Fig.~\ref{fig:extinction}(f) and, together with the full map and example spectra, Fig.~\ref{fig:spin_contrast_map}.

With $\{g,\kappa,\gamma_\mathrm{tot}\}/2\pi = \{0.84,\,13.5,\,0.052\}$\,GHz as quoted in the main text ($C=4.0$, $\mathcal{C}=0.96$), the optimum at $I_\mathrm{coil}=9$\,A ($\Delta\nu_\mathrm{spin}=308$\,MHz) occurs at $|\omega_{a,\downarrow}-\omega_c|\approx-2.3$\,GHz, giving $\mathcal{C}_\mathrm{spin}=0.91$ versus $0.81$ on resonance. The optimal detuning approaches $\delta\to0$ as $\Delta\nu_\mathrm{spin}\to\infty$ [Fig.~\ref{fig:spin_contrast_map}(a)], where the transitions become fully resolved and $\mathcal{C}_\mathrm{spin}$ recovers the cooperativity limit.

\section{Expected Zeeman splitting for given magnetic field} 
\label{app:expected_zeeman}
The magnetic field at the sample position is estimated from the coil geometry alone, with the coil--sample distance representing the dominant source of uncertainty.
At $I_\mathrm{coil} = \qty{9}{A}$ -- limited by superconducting quenching of the coil, observed at approximately \qty{9.3}{A} -- this yields $B_z = \qtyrange{0.10}{0.15}{T}$ and an expected optical Zeeman splitting of \qtyrange{350}{530}{MHz} for an unstrained SnV center.
The measured splitting of \SI{308\pm12}{\mega\hertz} is reproduced for a strain amplitude of $\Upsilon_g/2\pi = \qtyrange{110}{190}{GHz}$, assuming an angle of \ang{54.7} between the applied magnetic field and the SnV symmetry axis~\cite{rosenthal_microwave_2023,karapatzakis_microwave_2024}.
This is consistent with the observed \qty{39}{GHz} red-shift of the SnV transition frequency relative to the unstrained value $f^0_\mathrm{SnV} = \qty{484.134}{THz}$, as well as with the pronounced strain-induced polarization splitting of the cavity mode at the emitter position.

\section{Summary of system parameters}
\label{sec:system_parameters}

Table~\ref{tab:system_parameters} collects the cavity geometry, the expected Purcell enhancement for air-like and diamond-like hybrid cavity modes, and the measured cavity-QED parameters, together with how each value was obtained. The expected Purcell factor is estimated purely from the measured finesse $\mathcal{F}$ and derived mode waist $w_0$, with the uncertainties reflected through a Monte Carlo simulation. The cavity--emitter coupling itself is instead characterized directly from the resonant transmission measurements.

\begin{table*}[t]
  \centering
  \caption{System parameters.}
  \label{tab:system_parameters}
  \begin{tabular*}{\textwidth}{@{\extracolsep{\fill}} l l l @{}}
    \toprule
    Parameter & Value & Source\\
    \midrule
    \multicolumn{3}{@{}l}{\textit{Geometry}}\\
    Mirror radius of curvature $\mathrm{ROC}$& \SI{30}{\micro\meter} & measured\\
    Cavity air gap $t_\mathrm{a}$ & \qtyrange{1.0}{2.0}{\micro\meter} & estimate \\
    Diamond thickness $t_\mathrm{d}$ & \qtyrange{0.8}{2.1}{\micro\meter} & measured \\
    Cavity mode waist $w_{0,\mathrm{d}}$
      & \SI{1.2 \pm 0.1}{\micro\meter} & derived using $(\mathrm{ROC},t_a,t_d)$~\cite{dam_optimal_2018}\\
    \midrule
    \multicolumn{3}{@{}l}{\textit{Expected Purcell enhancement (air-like mode)}}\\
    Cavity finesse $\mathcal{F}$ & \num{4840 \pm 250} & measured [Fig.~\ref{fig:hybrid}(b)]\\
    Effective cavity length $L_\mathrm{eff}^\mathrm{air}$
      & \SI{3.7 \pm 0.1}{\micro\meter} & transfer-matrix simulation\\
    Mode volume $V^\mathrm{air}$
      & \SI{18 \pm 9}{\lambda^3} & transfer-matrix simulation\\
    Expected Purcell factor $\mathcal{F}_P$
      & \num{11 \pm 1} & Monte Carlo using $(\mathcal{F},w_0)$\\
    Expected effective Purcell factor $C_0$
      & \num{4.0 \pm 0.4} & Monte Carlo using $(\mathcal{F},w_0,\beta_\mathrm{tot})$\\
    \midrule
    \multicolumn{3}{@{}l}{\textit{Expected Purcell enhancement (diamond-like mode)}}\\
    Cavity finesse $\mathcal{F}$ & \num{1700 \pm 180} & measured [Fig.~\ref{fig:hybrid}(b)]\\
    Effective cavity length $L_\mathrm{eff}^\mathrm{dia}$
      & \SI{2.0 \pm 0.9}{\micro\meter} & transfer-matrix simulation\\
    Mode volume $V^\mathrm{dia}$
      & \SI{10 \pm 6}{\lambda^3} & transfer-matrix simulation\\
    Expected Purcell factor $\mathcal{F}_P$
      & \num{23 \pm 3} & Monte Carlo using $(\mathcal{F},w_0)$\\
    Expected effective Purcell factor $C_0$
      & \num{8.3 \pm 1.1} & Monte Carlo using $(\mathcal{F},w_0,\beta_\mathrm{tot})$ \\
    \midrule
    \multicolumn{3}{@{}l}{\textit{Cavity QED (measured)}}\\
    SnV transition frequency $f_0$ & \SI{484.095}{\tera\hertz} & fit of Fig.~\ref{fig:extinction}(b)\\
    Free-space lifetime $\bar{\tau}_0$ & \SI{6.1 \pm 0.4}{\nano\second} & Fig.~\ref{fig:purcell}(e)\\
    Lifetime-limited linewidth $\gamma_0/2\pi$ & \SI{26.1 \pm 0.2}{\mega\hertz} & derived from $\bar{\tau}_0$\\
    Total emitter linewidth $\gamma_\mathrm{tot}/2\pi$ & \SI{51.5 \pm 16}{\mega\hertz} & fit of Fig.~\ref{fig:extinction}(b)\\
    Single-photon Rabi frequency $g/2\pi$ & \SI{0.84 \pm 0.02}{\giga\hertz} & fit of  Fig.~\ref{fig:extinction}(b)\\
    Cavity linewidth $\kappa/2\pi$ & \SI{13.5 \pm 0.3}{\giga\hertz} & fit of Fig.~\ref{fig:extinction}(b)\\
    Cavity quality factor $Q$ & \SI{35900\pm 800}{} & derived from $(\kappa, f_0)$\\
    Measured cooperativity $C$ & \num{4.0 \pm 1.3} & derived from $(g,\kappa,\gamma_\mathrm{tot})$\\
    Ideal cooperativity $C_0$ & \num{8.0 \pm 0.7} & derived from $(g,\kappa,\gamma_0)$\\
    Resonant extinction contrast $\mathcal{C}$ & \num{0.96 \pm 0.02} & derived from $C$\\
    \bottomrule
  \end{tabular*}
\end{table*}

\section{Resonant extinction of a second SnV center}

\begin{figure*}[t]
    \centering
    \includegraphics[width=\textwidth]{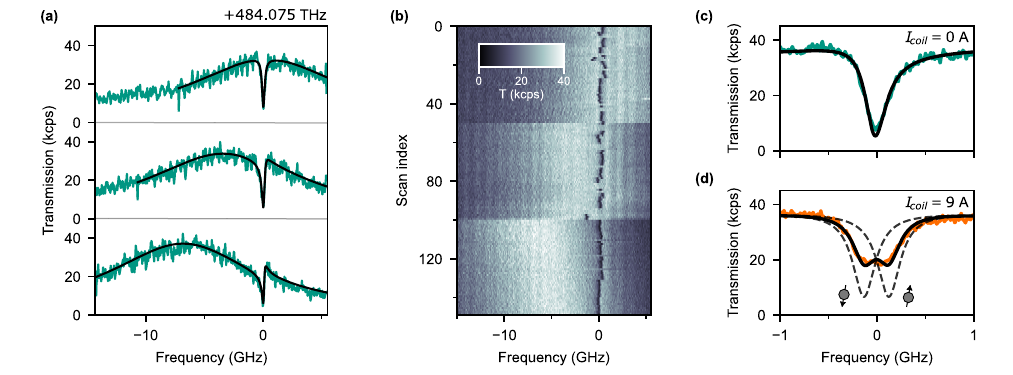}
    \caption{Resonant extinction spectroscopy of a second SnV center.
    (a)~Example transmission traces (green) together with fits to the cQED model (black) at three different cavity--emitter detunings, illustrating the on-resonance extinction dip and the Fano-like asymmetric lineshape appearing away from resonance.
    (b)~Corresponding full set of transmission scans versus frequency and scan index at zero magnetic field, combining the three detuning conditions shown in~(a); the narrow extinction feature shows pronounced spectral wandering between consecutive scans.
    (c)~Transmission spectrum averaged near zero cavity--emitter detuning at zero coil current (teal) together with the cQED fit (black), from which the coupling parameters $\{g,\kappa,\gamma_\mathrm{tot}\}/2\pi$ quoted in the text are extracted.
    (d)~Averaged transmission spectrum at a coil current of $I_\mathrm{coil}=9$\,A (orange), showing the magnetic-field-induced splitting of the two spin-conserving transitions; the black solid line is a fit to the split cQED model of Eq.~\eqref{eq:split_cqed}, decomposed into its two branches (black dashed).}
    \label{fig:second_emitter}
\end{figure*}

A second SnV center, coupled to a similar diamond-like cavity mode, was characterized via resonant extinction spectroscopy. It is not featured in the main text because it exhibited worse charge stability and larger pure dephasing, which lowers the coherent cooperativity. Individual example traces at three cavity--emitter detunings, shown in Fig.~\ref{fig:second_emitter}(a), display the on-resonance extinction dip together with the Fano-like asymmetric lineshape observed away from resonance; the full set of corresponding transmission scans at zero magnetic field is shown in Fig.~\ref{fig:second_emitter}(b). The spectrum averaged near zero cavity--emitter detuning, shown in Fig.~\ref{fig:second_emitter}(c), yields the cQED parameter set
\begin{equation*}
    \{g, \kappa, \gamma_\mathrm{tot}\}/2\pi = \{0.81(4),\, 16.2(5),\, 0.12(2)\}\,\text{GHz},
\end{equation*}
from which we obtain a coherent cooperativity of $C=1.4(4)$ and a resonant extinction contrast of $\mathcal{C}=0.81(5)$. The coherent coupling rates agree, within the uncertainty, with those of the SnV center discussed in the main text; the cooperativity is significantly lower here because the pure dephasing rate is a factor of 4--5 above the transform-limited linewidth. This is also apparent in the individual scans of Fig.~\ref{fig:second_emitter}(b), which show markedly larger spectral jumps between consecutive scans.

Applying a magnetic field with a coil current of $I_\mathrm{coil}=\SI{9}{\ampere}$ lifts the degeneracy of the two spin-conserving transitions, as shown in Fig.~\ref{fig:second_emitter}(d). Fitting the averaged data with the split cQED model of Eq.~\eqref{eq:split_cqed} yields a splitting of $\delta\nu_\mathrm{spin}=\SI{263\pm16}{\mega\hertz}$, significantly smaller than the value obtained in the main text. This is consistent with larger strain at this sample position, indicated by the further red-shifted ZPL frequency of $f_0=\SI{484.075}{\tera\hertz}$, and with an increased susceptibility to fluctuations of the charge environment, which manifests as larger pure dephasing and spectral diffusion. As a consequence of a reduced splitting and larger pure dephasing, both spin transitions overlap more, which together with a reduced cooperativity leads to a spin contrast of only $\mathcal{C}_\mathrm{spin}=0.64$, which highlights the importance of near transform-limited emitters.